\documentclass[12pt,preprint]{aastex}
\def\paren#1{\left( #1 \right)}
\def\bra#1{\left[ #1 \right]}
\def\angle#1{\left\langle #1 \right\rangle}
\def\ltsima{$\; \buildrel < \over \sim \;$}
\def\lsim{\lower.5ex\hbox{\ltsima}}
\def\gtsima{$\; \buildrel > \over \sim \;$}
\def\gsim{\lower.5ex\hbox{\gtsima}}
\newcommand{\bhat}[1]{{\hat {\bf #1}}}
\usepackage{graphicx}
\usepackage{epstopdf}
\usepackage{amsmath}
\begin{document}
\title{Ejection and Capture Dynamics in Restricted Three-Body
Encounters}
\author{Shiho Kobayashi\altaffilmark{1}, 
        Yanir Hainick\altaffilmark{2,3},
        Re'em Sari\altaffilmark{3,4}, 
     and Elena M. Rossi\altaffilmark{5}}

\altaffiltext{1}{Astrophysics Research Institute, Liverpool John Moores
University, Birkenhead CH41 1LD, UK}
\altaffiltext{2}{Raymond and Beverly Sackler School of Physics and
Astronomy, Tel Aviv University, Tel Aviv 69978, Israel}
\altaffiltext{3}{Racah Institute of Physics, Hebrew University, Jerusalem 91904, Israel}
\altaffiltext{4}{Theoretical Astrophysics 350-17, California Institute
of Technology, CA 91125, USA}
\altaffiltext{5}{Leiden Observatory, Leiden University, PO Box 9513,
2300 RA Leiden, the Netherlands}

\begin{abstract}
 We study the tidal disruption of binaries by a massive point
 mass (e.g. the black hole at the Galactic center), and we discuss
 how the ejection and capture preference between unequal-mass binary
 members depends on which orbit they approach the massive object. 
 We show that the restricted three-body approximation provides a
 simple and clear description of the dynamics.  
 The orbit of a binary with mass $m$ around a massive object $M$
 should be almost parabolic with an eccentricity 
 $|1-e| \lsim (m/M)^{1/3} \ll 1$ for a member to be captured,
 while the other is ejected. Indeed, the energy change of the 
 members obtained for a parabolic orbit can be used to describe
 non-parabolic cases. If a binary has an encounter velocity much larger
 than $(M/m)^{1/3}$ times the binary rotation velocity, it would be
 abruptly disrupted, and the energy change at the encounter 
 can be evaluated in a simple disruption model.  
 We evaluate the probability distributions for the ejection and capture
 of circular binary members and for the final energies. 
In principle, for any hyperbolic (elliptic) orbit, the heavier member
 has more chance to be ejected (captured), because it carries a larger
 fraction of the orbital energy. However, if the orbital energy is close
 to zero, the difference between the two members becomes small, and
 there is practically no ejection and capture preference. 
 The preference becomes significant
 when the orbital energy is comparable to the
 typical energy change at the encounter. 
 We discuss its implications to hypervelocity stars and
 irregular satellites around giant planets. 
 \end{abstract}

\section{Introduction}

The disruption of a star by a massive black hole (BH) is one of the
most spectacular examples of the tidal phenomena (Komossa \& Bade
1999; Donley et al. 2002; Grupe et al. 1995). A star that wanders too
close to a massive BH is torn apart by gravitational forces. Almost half
the debris would escape on hyperbolic orbits, while the other half would
traverse elliptic orbits and returns to periapsis before producing
a conspicuous flare (e.g. Rees 1988). The disruption process has been
numerically investigated in detail (Evans \& Kochanek 1989; Laguna et
al. 1993; Ayal et al. 2000; Kobayashi et al. 2004; Guillochon et
al. 2009), and the new generation of all-sky surveys are expected to
detect many tidal flares (Strubbe \& Quataert 2009, Lodat \& Rossi
2011). Recently a possible discovery of the onset of the rapid BH
accretion has been reported (Burrows et al. 2011; Zauderer et al. 2011; 
Levan et al. 2011; Bloom et al. 2011).

Once a star gets deeply inside the tidal radius of a BH, the tidal
force dominates over the self-gravity and thermal pressure of the
star. A very simplified description of the disruption process could
be the encounter between a star cluster (or a cluster of point masses) and a
massive BH. The simplest case consists of a binary and a massive BH
in which after the tidal disruption, one star would escape to the infinity,
while the other could be captured by the BH. This is actually one of the
leading models for the formation of hypervelocity stars
(Hills 1988; Yu\& Tremaine 2003). The captured stars may explain the S-stars
in the Galactic center (Gould \& Quillen 2003; Ginsburg \& Loeb 2006;
Ghez et al. 2005; Genzel et al. 2010).

Hypervelocity stars are stars with a high velocity exceeding the escape
velocity of the Galaxy. After the discovery of such stars
in a survey of blue stars within the Galactic halo (Brown et
al. 2005; Hirsch et al. 2005; Edelmann et al. 2005),  many authors
have predicted the properties of the hypervelocity stars (Gualandris et
al. 2005; Bromley et al. 2006; Sesana et al. 2007; Peretz et al. 2007;
Kenyon et al. 2008; Tutukov \& Fedrova 2009; Antonini et al. 2010; Zhang
et al. 2011). These investigations so far have used three-body simulations
or analytic methods that relied on results from three-body simulations.

The six orders of magnitude mass ratio between the Galactic center BH
and the binary stars allows us to
formulate the problem in the restricted three-body approximation. In a
previous paper (Sari et al. 2010, hereafter SKR), we have shown that the
approximation is efficient and useful to understand
how binary stars behave at the tidal breakup when the binary's center of
mass approaches the BH in a parabolic orbit. In this paper, we
generalize the approximation for orbits with arbitrary
eccentricity. This enables us to give a complete picture 
of the ejection and capture process.
We also provide the ejection and capture probability distributions that
can be simply rescaled in terms of binary masses, their initial
separation and the binary-to-black hole mass ratio
when applied to a specific system. Our method is computationally more
efficient than full three-body simulations, and it is easier to grasp
the nature of the tidal interaction.

In  \S2, we outline the restricted three-body approximation. In \S3, we
evaluate how much energy each member gains or loses at the tidal
encounter and we discuss how the energy change evaluated for a parabolic
orbit can be used to study non-parabolic orbit cases, and in \S4, we give
qualitative discussion on the ejection and capture preferences.
In \S5, we study high velocity encounters.
In \S6, the numerical results are discussed. In \S7, we use our results to
describe the capture process of Triton around Neptune. Finally, in \S8,
we summarize the results.

\section{The Restricted Three-Body Problem}\label{sec:Restricted Three-Bobdy Problem}
The equation of motion for each of the binary members is given by
\begin{equation}
\ddot{\bf r}_1=-{G M \over r_1^3}{\bf r}_1 +
{Gm_2 \over |{\bf r}_1- {\bf r}_2|^3}  ({\bf r}_2- {\bf r}_1),
\end{equation}
\begin{equation}
\ddot{\bf r}_2=-{G M \over r_2^3}{\bf r}_2 -
{Gm_1 \over |{\bf r}_1-{\bf r}_2|^3} ({\bf r}_2- {\bf r}_1),
\end{equation}
where ${\bf r}_1$ and ${\bf r}_2$ are the respective distance from the
massive point mass with $M$. We will call the point mass the black hole
(BH), though the binary is assumed to travel well outside the event
horizon and our results can be applied to any systems which include a
Newtonian massive point mass. The equation for the distance
between the two ${\bf r}\equiv {\bf r}_2-{\bf r}_1$ is
\begin{equation}
\label{vecr}
\ddot {\bf r}=-{G M \over r_2^3} {\bf r}_2+
{G M \over r_1^3}{\bf r}_1 - {Gm \over r^3} {\bf r}.
\end{equation}
where $m=m_1+m_2 \ll M$.
We assume that the two masses are much closer to each other, and
to the trajectory of the center of mass of the binary ${\bf r}_{\rm m}$,
than each of them to the BH. 
Both energy and orbit obtained under the approximation are
fairly accurate except a part of the orbit just around the periapsis
passage (see SKR for the details).

Linearizing the first two terms of equation (\ref{vecr}) around the
center of mass orbit ${\bf r}_{\rm m}$, we find that the zero orders
cancel out. Then, rescaling the distance between the bodies by
$(m/M)^{1/3} r_{\rm p}$ and the time by $\sqrt{r_p^3/GM}$
where $r_p$ is the distance of the closest approach between 
the center of mass of the binary and the BH, 
we can re-write eq.~(\ref{vecr}) in terms of the {\it dimensionless}
variables: ${\boldsymbol \eta}\equiv  (M/m)^{1/3}  ({\bf r}/r_{\rm p})$ 
and $t$:
\begin{equation}
\ddot {\boldsymbol \eta}=\left( r_p  \over r_{\rm m}\right)^3
\left[ - {\boldsymbol \eta}+3 ({\boldsymbol \eta}
\bhat r_{\rm m})\bhat r_{\rm m} \right]
- { {\boldsymbol \eta} \over |{\boldsymbol \eta}|^3},
\label{eq:r_vec}
\end{equation}
where $\bhat r_{\rm m}$ is a unit vector pointing the center of mass
of the binary. We define the orbit of the center of mass to be a conic orbit
$r_{\rm m}/r_p=(1+e)/(1+ e \cos f)$
where $e$ is the eccentricity, and the true anomaly $f$ is the angle
from the point of closest approach. Since $\bhat r_{\rm m}=(\cos f, \sin f, 0)$,
and we set ${\boldsymbol \eta}=(x,y,z)$, explicit equations in terms of
dimensionless Cartesian coordinates reads
\begin{eqnarray}
\label{ddotx}
\ddot x &=&
\frac{(1+e \cos f)^3}{ (1+e)^3}
\left[ - x+3 (x \cos f+y \sin f )\cos f   \right]
- {x \over (x^2+y^2+z^2)^{3/2} }\,, \\
\label{ddoty}
\ddot y &=&
\frac{(1+e \cos f)^3}{ (1+e)^3}
\left[ -y+3 (x \cos f+y \sin f )\sin f \right]
  - {y \over (x^2+y^2+z^2)^{3/2} }\,, \\
\label{ddotz}
\ddot z &=& - \frac{(1+e \cos f)^3}{ (1+e)^3}
z   - {z \over (x^2+y^2+z^2)^{3/2} }\,.
\end{eqnarray}
where the eccentricity
\begin{equation}
e=1+\frac{2r_pE}{GMm}
\label{eccentricity}
\end{equation}
is related to the energy of the center of mass
which is given by
\begin{eqnarray}
E &=& \frac{m }{2}|\dot{\bf r}_{\rm m}|^2-\frac{GMm}{r_{\rm m}}.
\end{eqnarray}
Using the dimensionless time, the conservation of the angular
momentum can be expressed as
\begin{equation}
\dot{f}=\paren{1+e}^{-3/2}\paren{1+e\cos f}^2.
\label{dotf}
\end{equation}
Analytically one has relations between $r_{\rm m}$ and $t$ through a parameter
which are given by (e.g. Landau \& Lifshitz 1976) 
\begin{eqnarray}
&E<0& \ \ r_{\rm m}/r_p=(1-e)^{-1}\paren{1-e \cos \xi}, \ \
t=(1-e)^{-3/2} \paren{\xi - e \sin \xi}, \\
&E=0& \ \ r_{\rm m}/r_p=\paren{1+\xi^2}, \ \
t=\sqrt{2}  \paren{\xi+\xi^3/3}, \\
&E>0& \ \  r_{\rm m}/r_p= (e-1)^{-1} \paren{e \cosh \xi -1}, \ \
t=(e-1)^{-3/2} \paren{e \sinh \xi -\xi},
\end{eqnarray}
where the closest approach $r_{\rm m}=r_p$ happens at $t=0$.

\section{Energy change at the BH encounter}
We are interested in the fate of stars in a binary, following its
encounter with a massive BH. In order to study the ejection and capture
process, we evaluate the energies of the stars as functions of
time. When the binary is at a large distance from the BH, the binary
members rotate around their center of mass which gradually
accelerates towards the BH. The specific self-gravity energy of the
binary is about $-v_0^2\equiv-Gm/a$. Analytic arguments 
(SKR) suggest that at the tidal breakup one member gets additional energy of
the order of $v_{\rm m} v_0$ where $v_{\rm m}$ is the velocity of the
center of mass  at the tidal radius $r_t=(M/m)^{1/3}a$. If the
binary approaches the BH with negligible orbital energy, the velocity is
$v_{\rm m}=(GM/r_t)^{1/2}=v_0(M/m)^{1/3}$. The additional
energy is larger than the self-gravity energy by a factor of
$(M/m)^{1/3}\gg 1$. Therefore, we will neglect the self-gravity term in
the following energy estimates. This treatment is valid as long as the
binary is injected into the orbit ${\bf r}_{\rm m}$ at a radius 
much larger than the tidal radius.

The energy of one binary member $m_i$ is given by
\begin{equation}
E_i=\frac{m_i}{2}|\dot{\bf r}_i|^2-\frac{GMm_i}{r_i}.
\end{equation}
Linearizing the kinetic and potential energy terms around the orbit of
the center of mass ${\bf r}_m$ and using the initial energy 
$I_i\equiv(m_i/m)E$, we obtain
\begin{eqnarray}
E_i&=& I_i+ \Delta E_i, \\
\Delta E_i &\equiv& m_i \dot{\bf r}_m\paren{\dot{\bf r}_i-\dot{\bf r}_m}
+\frac{GMm_i}{r_m^3}{\bf r}_m\paren{{\bf r}_i-{\bf r}_m},
\end{eqnarray}
Since in our limit the total energy of the system is $E$, considering
$\Delta E_2=-\Delta E_1$, we get
\begin{equation}
\Delta E_2= \frac{m_1m_2}{m}
\paren{\dot{\bf r}_m \dot{\bf r}
+\frac{GM}{r_m^3}{\bf r}_m {\bf r}}.
\label{E2}
\end{equation}
Using our rescaled variables, the additional energy is given by
\begin{eqnarray}
\Delta E_2 &=& -\Delta E_1=
\frac{Gm_1m_2}{a}\paren{\frac{M}{m}}^{1/3} \Delta \bar{E},
\label{DEandF}
\\
\Delta \bar{E} &\equiv& D^{-1}
\bra{
\frac{-\dot{x}\sin f+\dot{y}(e+\cos f)}{\sqrt{1+e}}
+\frac{(1+e\cos f)^2}{(1+e)^2}\paren{x\cos f+y\sin f}},
\label{dimlessdE}
\end{eqnarray}
where $D= r_p/r_t$ is the penetration factor which is useful 
to characterize the tidal encounter.
Once the binary dissolves, $\Delta \bar{E}$ becomes
a constant because the body is eventually moving only under the
conservative force of the BH. Hereafter, the energy change $\Delta
\bar{E}$ means the constant value after the disruption,  otherwise we
specify it. The equation of motion
(\ref{eq:r_vec}) indicates that the negative of a solution ${\bf r}={\bf
r}(t)$ is also a solution. The energies $\Delta E_i$ are also linear in the
coordinates. Therefore, another binary starting with a phase difference $\pi$
will have the same additional energy in absolute value but opposite in
sign. A uniform distribution in the binary phase implies that, when the
binary is disrupted, each body has a $50\%$ chance of gaining
energy (and a $50\%$ chance of losing energy).

As we have discussed, the typical energy change is larger than the
self-gravity energy by a factor of $\sim (M/m)^{1/3}$,
it is of order of $(Gm_1m_2/a)(M/m)^{1/3}$.
Then, the dimensionless quantity $\Delta \bar{E}$ is an order-of-unity
constant after the disruption. Its exact value depends on orbital parameters,
but for qualitative discussion we just need to know that
$\Delta \bar{E}$ is about unity. Later we will numerically show that
$\Delta \bar{E}$ is an order-of-unity in the relevant parameter 
regime\footnote{For prograde orbits with $D\sim 0.1$, the energy change is as
large as $\Delta \bar{E} \sim 30$ in a very narrow range of the binary
phase (see figure 7 in SKR) where the binary members once come close to 
each other before they break up.  However, the phase-averaged value $\langle
|\Delta \bar{E}| \rangle$ is still an order-of-unity and it is a more
relevant quantity for the discussion on the ejection and capture
probabilities and the final energies}, and numerical values will be used
to estimate the ejection and capture probabilities. 

Rescaling energies by the typical value of the energy change, the
energies of the binary members after the disruption are given by
\begin{equation}
\label{Ei}
\bar{E}_1 =  \bar{I}_1-\Delta \bar{E},   \ \
\bar{E}_2 =  \bar{I}_2+\Delta \bar{E},
\end{equation}
where bar denotes energy scaled by $(Gm_1m_2/a) (M/m)^{1/3}$.
An interesting outcome of the encounter between a binary system and a
massive BH is the ``three-body exchange reaction'' 
(Heggie 1975; Hills 1975) where one member of the
binary is expelled and its place is taken by the BH, i.e. one binary member
is captured by the BH and the other is ejected to infinity. In order for a
member $m_i$ to escape from the BH,
the initial binding energy should be smaller than the energy gain:
$|\bar{I}_i| < |\Delta \bar{E}| \sim 1$. The same condition is
required when a member of the binary in a hyperbolic orbit loses energy
and it is bound around the BH. Therefore, when we discuss the
ejection or capture process associated with a massive BH, the absolute
value of the initial energy should be comparable or less than unit:
$|\bar{I}_i|\lsim 1$.

Since the energy, penetration factor (periapsis radius) and eccentricity
are related by eq (\ref{eccentricity}) or equivalently:
\begin{equation}
e=1+2D\bar{E}\paren{\frac{m}{M}}^{1/3} \paren{\frac{m_1}{m}}
 \paren{\frac{m_2}{m}},
\label{eq:ecc-energy}
\end{equation}
only two of them are the independent parameters to describe the binary orbit.
Considering $|\bar{I}_i|\lsim 1$ together with the mass ratio
$(m/M)^{1/3}\ll 1$ and the tidal disruption condition $D\lsim 1$, 
the eccentricity should be almost unity 
$|1-e| \lsim D (m/M)^{1/3} (m_{par}/m)$ for a member $m_i$ to be ejected
or captured where $m_{par}$ is the mass of the partner ($m_{par}=m_2$
for $i=1$ and $m_1$ for $i=2$). If we use the semi-major axis 
$r_a \equiv r_tD/(1-e)$, the condition can be rewritten as  $|r_a/r_t|\gsim
(M/m)^{1/3} (m/m_{par})$ where $r_a$ is negative for hyperbolic orbits. 

Such orbits differ very little from parabolic orbits with the same
periapsis distance, especially around the tidal radius and inside it.
Therefore, the energy change $\Delta \bar{E}$ is expected to be
almost identical to that for the parabolic case. As long as we study the
exchange reaction, we can approximate $\Delta \bar{E}$ by the
parabolic results $\Delta \bar{E}_{e=1}$. However, $e\sim 1$ does not
necessarily mean $|\bar{E}|\ll 1$. In general, we need to take into
account the offset of the final energy due to the non-zero initial
energy, which would affect the ejection and capture probabilities. The
final energies are approximately given by 
\begin{equation}
\label{Eipara}
\bar{E}_1 =  \bar{I}_1-\Delta \bar{E}_{e=1},   \ \
\bar{E}_2 =  \bar{I}_2+\Delta \bar{E}_{e=1}.
\end{equation}

\section{Which gets kicked out?}\label{sec:Which gets kicked out?}

We here consider a simple question: Which member is ejected or captured
if an unequal-mass binary is tidally disrupted by a massive BH? 
If $\bar{E}>0$ (hyperbolic orbits), one binary member could lose energy
and get captured by the BH, the other flies away with a larger
energy. Assuming a uniform distribution in the binary phase, each member
has a $50\%$ chance of losing energy (and gaining energy). 
However, since the lighter one (the secondary) has a smaller initial
energy, it is preferentially captured and the heavier one (the primary)
has more chance to be ejected. 

For elliptical orbits, by considering a plausible semi-major axis $r_a$,
we can obtain tighter constraints on the eccentricity and energy,
 compared to the requirements from the exchange reaction.
This is particularly relevant for studies of hypervelocity stars.
If $r_a$ is around the radius of influence of the BH
$r_{\rm h} \sim GM/\sigma^2$ where $\sigma$ is the local stellar
velocity dispersion, 
for the Galactic center, it is about $r_a \sim$ a few
parsecs $\sim 10^5 r_t$ for $a \sim$ several solar radii. Then, 
we get $1-e = D(r_t/r_a) \lsim 10^{-5}$ and  $|\bar{E}| \sim
(r_t/r_a) (M/m)^{1/3}(m/m_1)(m/m_2)\sim  10^{-3}$.
Our previous estimates based on parabolic orbits are appropriate
to study the production of hypervelocity stars for which an equal
ejection chance is expected (SKR). When the semi-major axis is as small as
$r_a \sim (M/m)^{1/3} (m/m_{par})r_t$, 
the initial energy $|\bar{I}_i|$ 
would be of the order of unity as we have discussed and it affects the 
ejection preference. Since the secondary has less negative initial
energy, it is preferentially ejected. 

Recently,  Antonini et al. (2011) performed N-body simulations of
unequal mass binaries with $m_1=6 M_\odot$, $m_2=1$ or $3 M_\odot$ 
and $a=0.1$AU in elliptical orbits around a supermassive
BH $M=4\times10^6 M_\odot$. They find that the initial distance of the
binary from the central BH plays a fundamental role in determining which
member is ejected: for a large initial distance $d=0.1$pc or 
equivalently $r_a \sim 3\times 10^3 r_t$, the ejection probability is
almost independent on the stellar mass,  while for $d=0.01$pc or 
$r_a \sim 3\times 10^2 r_t$, the lighter star is 
preferentially ejected. Considering that the ejection probability 
significantly decreases if $r_a$ becomes smaller than 
$\sim (M/m)^{1/3} (m/m_{par})r_t \sim 
80 (m/m_{par}) r_t$, these results are consistent with our analysis.   

These ejection preferences for hyperbolic and elliptic orbits are
naturally understood if we consider a 
large mass ratio for the binary members. The energy of the primary
practically does not change at the tidal encounter.  Whether it is
ejected or captured after the tidal breakup simply depends on the
initial energy $\sim E$, while the secondary might have a chance to make
a transition between a bound and unbound orbits around the BH 
(Bromley et al. 2006). In the large mass ratio limit, the exchange
reaction condition (i.e. the transition condition for the
secondary) is $|E| \lsim (Gm^2/a)(M/m)^{1/3}$ 
or equivalently $|r_a| \gsim (M/m)^{2/3}a$.

\section{High Energy Regime}\label{sec:High}
If a binary has a large orbital energy $\bar{E}\gg 1$, both members
are ejected after the BH encounter as a binary system or two
independent objects  
\footnote { 
If $\bar{E}$ is a large negative value, both members are captured after 
the disruption. Since the velocity at the tidal radius is reduced 
$v_m \lsim (M/m)^{1/3}v_0$, the energy change should be smaller
$|\Delta \bar{E}|\lsim 1$.}. 
Although the high energy regime is not important in the context of the
three-body exchange reaction, we discuss the regime to clarify the parameter
dependence of the numerical results in the next section. A high 
orbital energy $\bar{E}\gg (M/m)^{1/3}(m/m_1)(m/m_2)$
affects the velocity of their center of mass at the encounter 
$v_m \sim
(E/m+GM/r_m)^{1/2}=(M/m)^{1/3}v_0\sqrt{(e-1)/2D+r_t/r_m}$.
Then, the tidal disruption radius (i.e. where a binary is disrupted) can
be defined in three different ways. We here order them from a
large to small radius. 
(a) {\it Relative acceleration:} the radius at which the BH tidal force
becomes comparable to the mutual gravity of the binary. This is $r_t$. 
(b) {\it Relative velocity:} the radius at which the tidal force induces 
the relative velocity between the binary members comparable to the
binary escape velocity $v_0$.  
(c) {\it Relative position:} the radius at which the difference in
position increases by more than the initial binary separation.  

The duration that the center of mass is 
around $r_m$ is of order of $\Delta t \sim r_m/v_m$. During this period,
the tidal acceleration of the relative motion of the binary members by
the BH is of order of $A\sim GMa/r_m^3$. The two radii (b) and 
(c) can be estimated from two conditions: 
$\Delta v = A\Delta t \sim v_0$ and
$\Delta x = A\Delta t^2 \sim a$, provided that the duration of the 
encounter is comparable or shorter than the binary rotation time-scale:
$\Delta t \lsim a/v_0$. If the energy is 
high $(e-1)/D=2\bar{E}(m/M)^{1/3}(m_1/m)(m_2/m)\gg1$, these conditions 
give $r_m=r_t D^{1/4}/(e-1)^{1/4}$  and  $r_t D/(e-1)$, respectively.
Since they should be larger than the periapsis distance, only the cases 
that satisfy $D\lsim (e-1)^{-1/3}$ for the radius (b) or $e \lsim 2$ 
for the radius (c) lead to the disruption. 
The radius (c) is basically the place at which the orbit of the center 
of mass makes its turn (i.e. the periapsis). If the energy is low
$(e-1)/D \ll 1$, all the estimates give the original tidal radius $r_t$. 

When we discuss the energy change $\Delta E$ at the tidal encounter, 
there are two important points which we should emphasize. First, 
the energy of each of the binary members in the BH frame changes
only due to the mutual force between the binary members. Secondly 
most of the work done by one member on the other, which is $\Delta E$, 
is done outside the tidal radius $r_t$.
The mutual force is of order of $Gm_1m_2/a^2$. During the binary
rotation time-scale $a/v_0$, the force acts over a length $\sim
(v_m/v_0)a$ in the BH frame. 
Therefore, the work is $W \sim (m_1m_2/m)v_mv_0$. Since the direction 
of the mutual force changes with the binary rotation, $\Delta E(t)$ 
oscillates with the amplitude of $W$. When the binary is disrupted,
$\Delta E$ becomes a constant value which is basically determined 
by the binary phase at the disruption. Then, we might expect that 
the final value of 
$\Delta E$ is a sinusoidal function of the binary phase for circular
binaries. As we will see later, this is actually the case for the high
energy encounters. Even with the largest estimate of the tidal radius 
(i.e. $r_t$), the duration of the encounter $\Delta t\sim r_t/v_m$ 
is shorter than the binary rotation time-scale by a factor of 
$\sim \sqrt{D/(e-1)}\ll 1$. The work during the encounter is negligible 
compared to the work $W$ which has been done outside $r_t$. On the 
other hand,  in the low energy regime, the duration of the encounter 
is comparable to the binary rotation time-scale. Considering that 
at the encounter the orbits of the members in the comoving frame of 
their center of mass should be significantly deformed from the 
original orbits (e.g. circular orbits) before they finally break up, 
the work during the encounter could induce deviation of $\Delta E(\phi)$ 
from a simple sinusoidal function. However, the typical value 
is still expected to be about $\Delta E \sim (m_1m_2/m) v_0 v_m$. 

In both the low and high energy regime, the typical energy change is 
given in a dimensionless form by 
\begin{equation}
\Delta \bar{E}\sim
\sqrt{1+\bar{E}\paren{\frac{m}{M}}^{1/3}\paren{\frac{m_1}{m}}\paren{\frac{m_2}{m}}}
=\sqrt{1+\frac{e-1}{2D}}=\sqrt{1-\frac{r_t}{2r_a}},
\label{DEgene}
\end{equation}
where we have assumed $r_m=r_t$ to estimate $v_m$. In the high energy
regime, the disruption might happen at a smaller radius, but $v_m$ is
determined by the orbital energy and it is insensitive to the choice 
of $r_m$. When $(e-1)/D\gg1$, the energy change becomes much larger than
unity. However, the energy gain is not significant compared to the
original energy $\bar{E}$, and one finds that the tidal encounter is not
an efficient acceleration process anymore.  

In the high energy regime, the energy change eq (\ref{dimlessdE})
can be evaluated by assuming that a binary is abruptly disrupted at 
the tidal radius $r_t$, since the work during the tidal encounter 
is negligible. For a circular coplanar binary:
$(x,y)=D^{-1} (\cos\phi_t,\sin\phi_t)$ and 
$(\dot{x},\dot{y})=\pm D^{1/2}(-\sin\phi_t,\cos\phi_t)$, we obtain
\begin{equation}
\Delta\bar{E}=\paren{1\pm\frac{1}{\sqrt{D(1+e)}}}\sin f_t \sin\phi_t
+\paren{1\pm\frac{1+e/\cos f_t}{\sqrt{D(1+e)}}}\cos f_t \cos\phi_t
\label{eq:sudden}
\end{equation}
where $\phi_t$ is the binary phase at the tidal radius,
$f_t$ is the negative value solution of $1+e\cos f_t=(1+e)D$ and the 
signature indicates a prograde (+) or retrograde (-) orbit. This 
is a sinusoidal function of the binary phase as we expected, and the square of
its amplitude is $3-r_t/r_a\pm2\sqrt{(1+e)D}$ which is larger
for prograde orbits and the difference between prograde and retrograde
orbits becomes smaller for deep penetrators $D\ll 1$, because
in this limit the binary center of mass approaches the BH in an almost
radial fashion.

If the disruption is abrupt, the
ejection and capture preference could be roughly illustrated in terms of
velocity (e.g Morbidelli 2006; Agnor \& Hamilton 2006). The binary
members rotate around their center of mass, such that their own motion
is half of time with and half of time against, the motion of the center
of mass $\dot{\bf r}_m$. The net velocity of the members relative to the
BH is accordingly increased or reduced. Since the secondary has a higher
rotation velocity, it has more chance that the net velocity exceeds 
or drops below the escape velocity from the BH. Then, it is
preferentially ejected to infinity or captured in a bound
orbit. However, for the full discussion of the process, we also need to
take into account the variation in the escape velocity or the variation
in the potential. The displacement of order $a$ in the position of each
member of the binary, at a distance of about $r_t$ from the BH, results
in a change in gravitational energy of $GMa/r_t^2 \sim v_0^2(M/m)^{1/3}$, 
this is comparable to the variation in the kinetic energy.
As we have done, it should be easier to discuss the overall effect
in the energy domain. In our formula, the energy change (\ref{E2})
includes both the variation of kinetic energy and potential energy.
For prograde orbits, the kinetic and potential terms 
cooperate and the net energy change is larger, the member on the
``outside track'' is expected to be ejected and its partner is captured
(the ``outside track'' could be well defined especially when the orbital
energy is large because the duration of the encounter 
is much shorter than the binary rotation time-scale).
On the other hand, for retrograde orbits the variation in the
gravitational energy would counteract that in the kinetic energy.  

\section{Numerical Results}
In this paper we focus on results for circular coplanar binaries,
though our formulae can be used to study the evolution of a binary with
arbitrary orbital parameters.  The orbit of a binary is assumed to be
initially circular in the comoving frame of the binary center of 
mass. The center of mass of the binary is in a prograde or retrograde
orbit around the BH (see SKR for the details of the numerical setup). 

For $M/m\gg 1$ the problem can be reduced to the motion of a single
particle in a time-dependent potential (``the restricted three-body
approximation'') described by the equations ~(\ref{ddotx})-(\ref{ddotz})
and (\ref{dotf}). The energy change, eq (\ref{dimlessdE}), depends only on
the penetration factor $D$, the eccentricity $e$, the initial binary
phase $\phi$, and the binary rotation direction. As we have shown,
when a binary member is captured by the BH and the other is ejected,
we can further reduce the number of the parameters by assuming $e=1$ to
approximate the additional energy. The effect of the eccentricity  
$e\ne 1$ is  taken into account through the non-zero initial orbital
energy of the center of mass. This method (\ref{Eipara}) will be called
``the parabolic  approximation''. For a large orbital energy 
$\bar{E}\gg (M/m)^{1/3}(m/m_1)(m/m_2)$,
``the sudden disruption approximation'' (\ref{eq:sudden})
would become valid, but the three-body exchange reaction does not take
place in this regime. In all the
numerical codes the time evolution of objects are evaluated by using a
fourth-order Runge-Kutta integration scheme.

\subsection{Energy change and probability distributions}
\begin{figure}[t!]
\includegraphics[height=18cm]{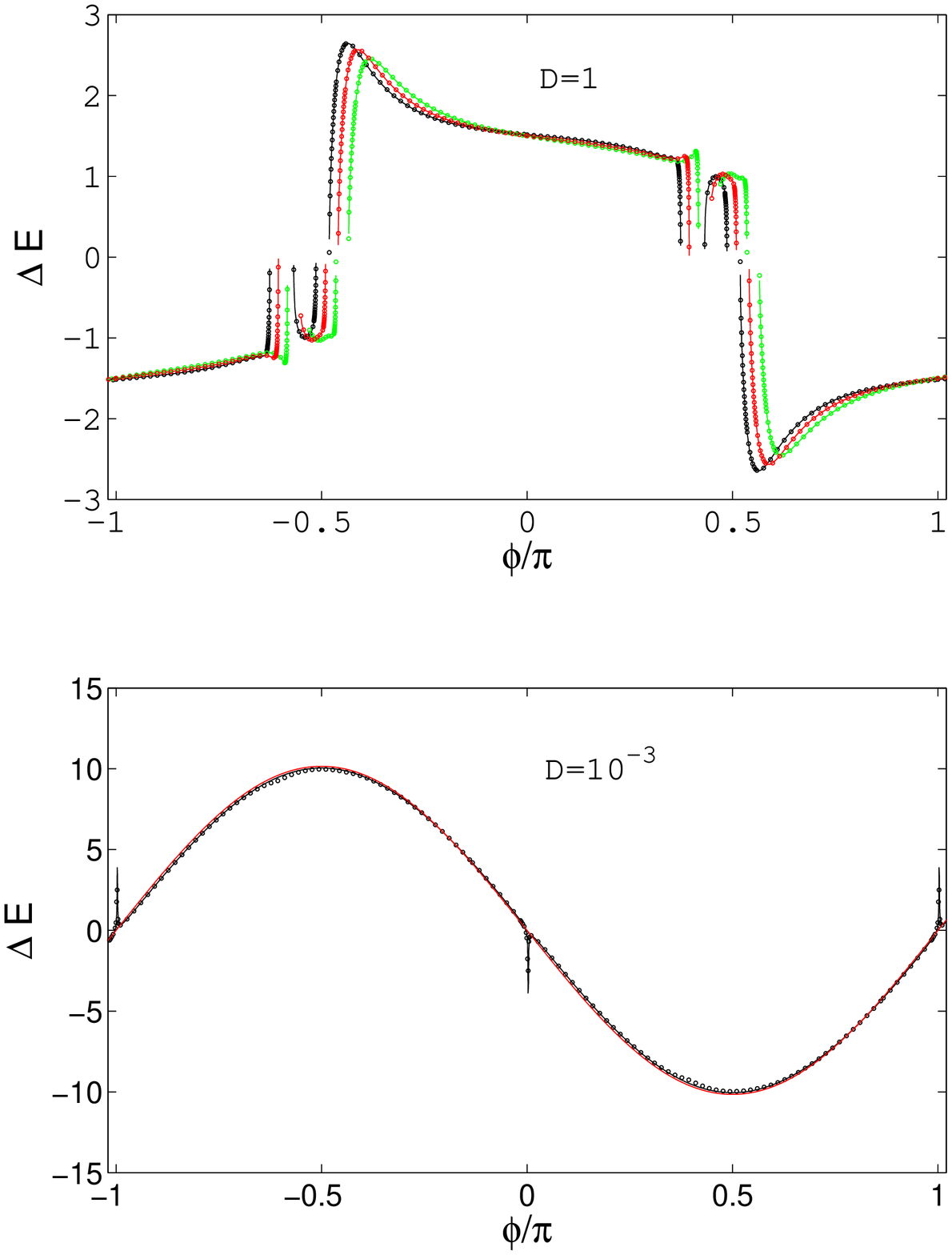}
\caption{
Energy change as a function of $\phi$:
top panel ($D=1$): the restricted three-body approximation
for $e=0.9$ (green solid line), $1$ (red solid line), and $1.1$ (black
 solid line). The full three-body calculations (circles).
Bottom panel ($D=10^{-3}$ and $e=1.1$): the restricted
three-body approximation (black solid line), the full three-body 
calculations (black circles) and the sudden disruption approximation 
(red solid line).
In the full three-body calculations, $M/m=10^6$ and
$m_1/m_2=3$ are assumed and the energy change is evaluated as 
$\Delta E=(m_1/m)E_2-(m_2/m)E_1$.
Prograde orbits are assumed for all the calculations.
Energy is in units of $(Gm_1m_2/a)(M/m)^{1/3}$.
\label{fig:DeltaE}}
\end{figure}

We have tested the restricted three-body approximation against the full
three-body simulations of a binary evolving around a massive object
(SKR). The full three-body orbit is accurately reproduced by the 
approximation equations, and the energy change, for example, differs at
a $0.1\%$ level when the exchange reaction happens. The comparison of
the energy change between the full three-body and restricted three-body
results is shown in figure \ref{fig:DeltaE}.  
Since they are in an excellent agreement, in the following discussion,
we will use the restricted three-body approximation to test the
parabolic and the sudden disruption approximation.

In figure \ref{fig:DeltaE}, one could notice the symmetry  
$\Delta \bar{E}(\phi+\pi)=- \Delta \bar{E}(\phi)$. The top panel 
shows that the energy change for $|e-1|/D=0.1$ is very similar to the parabolic
case with the same $D$ especially if we take into account the phase 
shifts\footnote{The initial distance of the binary center of mass to the
BH is assumed to be $r_0=15r_t$ for the parabolic calculations in the 
top panel of figure  \ref{fig:DeltaE}. As long as a simulation starts at a large
enough radius $r_0\gg r_t$, the results are largely independent of
it. However, a problem arises when we compare the phase dependence. 
If we assume the same initial radius for the non-parabolic
cases, it takes a slightly different time for the binary to reach the
vicinity of the BH, the binary interacts with the BH
with a slightly different binary phase. We have adjusted the initial
radii as the periapsis passage happens at the same time
($r_0\sim 12.2 r_t$ for $e=0.9$ and $\sim 17.4r_t$ for $e=1.1$). Since
this adjustment has been done neglecting the BH tidal field, the actual
phase at the periapsis is still different from the parabolic case.
This induces the phase shifts in the figure. In the bottom panel of
figure \ref{fig:DeltaE}, $r_0=15r_t$ is assumed for the restricted three-body
calculations, and in each case the binary phase is adjusted as the
results take the maximum value at $\phi=-\pi/2$.}. 
When the three-body exchange reaction happens, the value 
$|e-1|/D \lsim (m/M)^{1/3} (m_{par}/m) \ll 1$ should be very small,  
in such a case non-parabolic results tend toward a perfect overlapping
with the parabolic results.
In the bottom panel, the energy change in the high energy regime:
$(e-1)/D=100$ is shown. The sudden disruption approximation well
reproduces the three-body results, except for some notable spikes.

Most binaries are disrupted at the tidal encounter with a BH. However,
there are always finite phase regions where binaries survive (SKR). The
narrow gaps in the top panel of fig \ref{fig:DeltaE} correspond to such
regions. Although it is not evident in the bottom panel, very narrow
gaps exist just in the middle of the spikes. The high energy regime is
usually realized with a deep penetrator $D\ll 1$.  In such 
a case, we can ignore the self-gravity term ${\bf r}/r^3$ in eqs 
(\ref{ddotx})-(\ref{ddotz}), and free solutions are obtained. 
Actually, one of the free solutions: 
$(x,y) \propto (-\sin f, e+\cos f)$, which corresponds to the case that 
binary members have the same trajectory but are slightly separated in
time, dominates around the periapsis passage (SKR). 
Then, the binary is once disrupted at the tidal encounter, but 
after the periapsis passage at $t=0$, they come close to each other.
If we fine-tune the initial binary phase, they form a binary again. This 
produces the narrow gaps at the spikes. If the binary phase 
is slightly different from the fine-tuned values, they almost 
form a binary, but they eventually break up. The additional work 
at $t>0$ due to the mutual forces between the binary members produces 
the notable spikes.

\begin{figure}[t!]
\includegraphics[height=18cm]{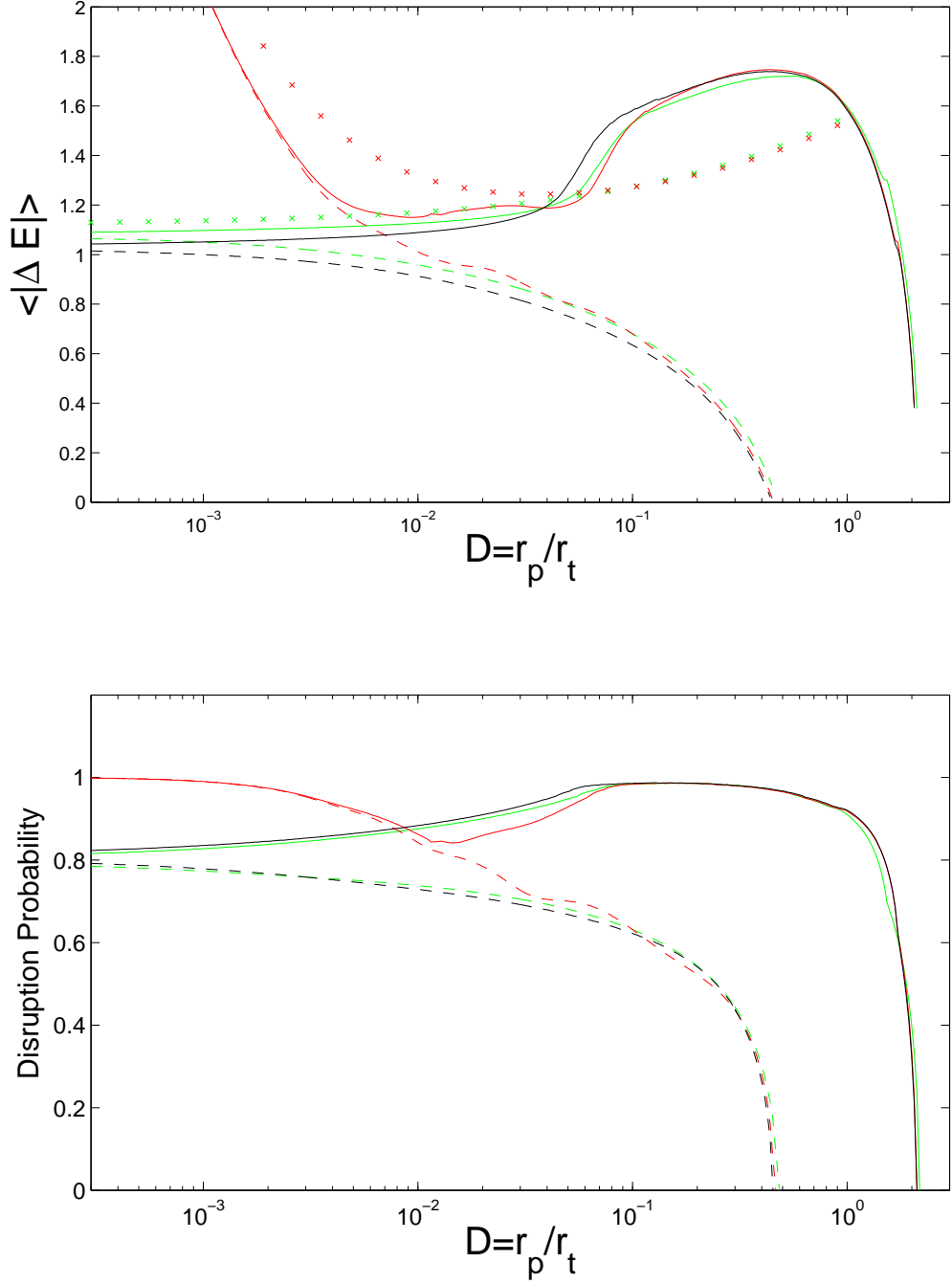}
\caption{
Phased-averaged energy change (top panel) and disruption probability
 (bottom panel) as a function of $D$. Results for the restricted three-body
 approximation are shown for $e=1$ (black lines),
 $e=1.01$ (red lines) and $(e-1)/D=0.1$ (green lines).
 Prograde (solid lines) and retrograde (dashed lines). 
 The sudden disruption approximation is shown for prograde orbits with 
 $e=1.01$ (red crosses) and $(e-1)/D=0.1$ (green crosses). 
 Energy is in units of $(Gm_1m_2/a)(M/m)^{1/3}$. 
\label{fig:D-dep}}
\end{figure}

The top panel of figure \ref{fig:D-dep} shows the phase-averaged
absolute value of the energy change $\angle{|\Delta \bar{E}|}$, 
as a function of the penetration factor $D$. The average is taken over 
the phase space where binaries are disrupted. We alternatively fix 
the orbital eccentricity or the orbital energy $\bar{E} \propto (e-1)/D$.
For a given eccentricity, a smaller $D$ corresponds to a larger 
orbital energy, while  for a given orbital energy, it corresponds to 
$e\to1$.  For $(e-1)/D \lsim 1$, the energy change remains of the order
of unity as we expect from eq (\ref{DEgene})
(see the black and green lines and the green crosses for the whole
disruption range, and the red lines and red crosses for $D \gsim 10^{-2}$). 
For $(e-1)/D \gsim 1$, instead, $\angle{|\Delta \bar{E}|}$ increases
towards smaller $D$ (the red lines and red crosses for $D \lsim 10^{-2}$).
Correspondingly, the disruption probability (bottom panel) 
becomes almost 100$\%$ for $(e-1)/D \gsim 1$. 
On the other hand, for the fixed energies 
$(e-1)/D=0$ and $0.1$, the disruption chance is about $80\%$ even in the
deep penetration limit $D\ll 1$ (the black and green lines).
Note that in the sudden disruption
approximation the disruption probability is 100\% by definition, and we
do not show it in the bottom panel. For our choice of $(e-1)/D=0.1$,
the behavior of $\angle{|\Delta \bar{E}|}$ always remains very similar 
to the parabolic orbit case, even when we use the sudden
disruption approximation especially for small $D$ (the green crosses in the
top panel). The same applies to the disruption probability function, but 
the sudden disruption approximation overestimates the disruption probability.
As it is well known, the retrograde binaries tend to be more stable
against the tidal encounter.   

For a given initial energy of a binary member $\bar{I}=(m_i/m)\bar{E}$ 
(in the following discussion we drop the subscript $i$ of the initial 
energy for simplicity), considering the symmetry of
$\Delta\bar{E}(\phi)$, the probability $P_{eje}(\bar{I})$ that the
member is ejected after the tidal encounter
is determined by the fraction of the binary phase region $[0, 2\pi]$ 
that satisfies $\Delta \bar{E}(\phi) < \bar{I}$, while the
capture probability  $P_{cap}(\bar{I})$ is determined by the fraction
satisfying $\Delta \bar{E}(\phi) > \bar{I}$. The sum is equal to the 
disruption probability: $P_{dis}=P_{eje}+P_{cap}$. 
Since $P_{eje}(-\bar{I})=P_{cap}(\bar{I})$, once we evaluate the
ejection probability $P_{eje}(\bar{I})$ for negative and zero energy, 
the distribution for positive energy 
$P_{eje}(\bar{I})=2P_{eje}(0)-P_{eje}(-\bar{I})$ and 
the capture probability for any energy
$P_{cap}(\bar{I})=P_{eje}(-\bar{I})$ are also obtained. 
The disruption probability is $P_{dis}=2P_{eje}(0)$.

We show in figure \ref{fig:peje} how the ejection probability of a binary 
member behaves for different penetration factors and binary orientations. 
Since the orbital energy and the eccentricity are not independent
parameters when $D$ is fixed: $e-1 \propto \bar{E}$, we need 
to assume a smaller eccentricity, in principle, for a larger 
negative orbital energy $\bar{I}$  (or larger negative $\bar{E}$) to estimate the
energy change $\Delta \bar{E}(\phi)$, the disruption probability 
and the ejection probability. However, the energy change and the
disruption probability are not so sensitive to the eccentricity, we
thus evaluate them by using parabolic orbit results (the parabolic
approximation) when the three-body exchange reaction happens. 
In the figure, the parabolic approximation calculations (the lines) are 
in a good agreement with the restricted three-body calculations (the circles).
For prograde orbits, the energy change $\Delta \bar{E} (\phi)$ 
is more sensitive to the phase $\phi$ around zero points, 
only very narrow phase regions satisfy $|\Delta \bar{E}|\lsim 1$
(see fig \ref{fig:DeltaE} top panel), then the ejection probabilities
have a plateau around $\bar{I}=0$.
The long low energy tail of the prograde $D=0.1$ case (the green solid line)
reflects the fact that the energy gain would be quite large for a narrow
binary phase region for this case.  The ejection (capture) probability
for $-\infty < \bar{I} < \infty$ is a monotonically increasing
(decreasing) function of the energy $\bar{I}$. The probabilities rapidly
change around $|\bar{I}| \sim 1$ and their function form depends 
mainly on $D$ and the binary rotation direction. 

\subsection{Probability distributions for various semi-major axes}

When we discuss an actual astrophysical system, it is more
physically intuitive to use the semi-major axis rather than the 
energy $\bar{I}$ or $\bar{E}$. By specifying the semi-major axis 
and the mass ratios, the ejection (or capture) probability
and the energy after the disruption can be evaluated. 
The orbital energy $\bar{E}$ is  given by 
\begin{eqnarray}
\bar{E}=-2.7 \paren{\frac{r_a/r_t}{100}}^{-1}
\paren{\frac{M/m}{10^6}}^{1/3}
\paren{\frac{m_1/m}{3/4}}^{-1}
\paren{\frac{m_2/m}{1/4}}^{-1}.
\label{eq:energy-ra}
\end{eqnarray}
$M/m=10^6$ and $m_1/m_2=3$  will be assumed in this section. 

Figure \ref{fig:ave} shows the characteristics of the ejection process
when a binary approaches the BH in elliptic orbits.  
To evaluate the ejection probability (the top panel), we have
not distinguished which member is ejected. Since both members are never 
ejected together when $\bar{E}< 0$, it is just the sum of the ejection
probabilities 
of the two members.  For parabolic orbits, one of the members is always
ejected if the binary is disrupted, then the black solid line indicates
the disruption probability. As $r_a/r_t$ becomes smaller, the
deviation of the orbit from the  parabolic one becomes larger. The
effect of the non-zero orbital energy is expected to become significant
around $r_a/r_t \sim (M/m)^{1/3} (m/m_2)\sim 400$, 
and the ejection probability rapidly decreases.  
The middle panel indicates which member in a binary is ejected more
frequently. There is no preference in the parabolic case. However,
the secondary is preferentially ejected for small $r_a/r_t$. 
The phase-averaged ejection energy is shown in the bottom panel. 
For parabolic orbits this is equivalent to the phase-averaged energy
change $\angle{|\Delta \bar{E}|}$. 

If we consider $P_{cap}(\bar{I})=P_{eje}(-\bar{I})$, it is possible to
interpret figure \ref{fig:ave} as the capture probability and capture preference
for the hyperbolic orbit cases with the same semi-major axises
in absolute value. The phase-averaged energy (the bottom panel) gives 
the absolute values of the averaged capture energies. 

\section{Irregular satellites around giant planets}

Over 150 satellites are orbiting the giant planets in the Solar system.
About one-third of these are classified as regular, with nearly 
circular and planar orbits. The majority of the satellites, however, 
are irregular ones which are more distant from their planet and
typically have larger eccentricities and/or inclination. Interestingly,
a large fraction of the irregular satellites orbit their planet in the
retrograde direction (Jewitt \& Haghighipour 2007).  
Triton, Neptune's largest moon, is among
them. Because of the retrograde orbit and composition similar to
Pluto's, Triton is thought to have been captured from the Kuiper
belt. Recently Agnor and Hamilton (2006) 
demonstrated that the gravitational encounter between Neptune and a
binary system, which had included Triton as a member, is an effective
mechanism to capture Triton. 
Since the mass ratio between Neptune 
$m_N\sim 17.2 M_\oplus$ and Triton $m_T\sim 3.58\times 10^{-3} M_\oplus$ 
is reasonably large $m_N/m_T\sim 4800$, we here revisit the capture process
from the point of view of the restricted three-body problem. 

We consider an encounter between Neptune and a binary system
with $m_1=m_T$, $m_2=0.1m_T$ and $a=20R_T$. The binary is assumed to 
be in a prograde hyperbolic orbit with the periapsis $r_p=8R_N$ where 
$R_T\sim 1.35\times10^3$ km and $R_N\sim 2.46\times10^4$ km are the radii of Triton
and Neptune, respectively. This set of the parameters are identical to 
what Agnor and Hamilton (2006) have assumed to obtain their figure 2. 
Since the tidal radius is about $r_t\sim 18 R_N$, the penetration factor
for the orbit is  $D\sim 0.45$.
For a given encounter velocity at infinity $v_\infty$, we evaluate the
binary disruption chance and the capture probability of the binary
members. The velocity at infinity $v_\infty = 1$ km/s
is related to the semi-major axis 
of the hyperbolic orbit as
\begin{eqnarray}
|r_a/r_t| &=&  16 \paren{\frac{v_\infty}{1 ~\mbox{km/s}}}^{-2}
\paren{\frac{a}{20R_T}}^{-1} 
\paren{\frac{m}{1.1 m_T}} 
\paren{\frac{M/m}{4.37 \times 10^3}}^{2/3},
\end{eqnarray}
which is about the critical value $|r_a/r_t|\gsim (M/m)^{1/3}\sim 16$
for the three-body exchange reaction. The numerical results are shown in
the figure \ref{fig:triton}, and the relevant case (the solid lines in
the top panel) is
in good agreement with the figure 2 in Agnor and Hamilton (2006). Since
the disruption chance depends mainly on $D$, for the fixed $D$,
it is constant $\sim 97\%$.
For high velocities, the probability for the primary drops sharply, while
it is still $\sim 50\%$ for the secondary because of its smaller initial
energy $\bar{I}$. If we assume a larger secondary mass $m_2=m_T/2$, the
energy change $\Delta E$ at the encounter would be larger. 
The $50\%$ capture probability of the primary now extends to 
higher velocities, while the capture rate of the secondary is similar
because the primary mass is fixed (the dashed lines). 
The black dashed-dotted lines indicate the capture probabilities 
obtained by using the parabolic approximation for $m_2=0.1m_T$. 
Even at the high velocity $v_\infty\sim 1.6$
km/s or equivalently high eccentricity
$e-1=D(m/M)^{2/3}(av_\infty^2/Gm)\sim 7 \times10^{-2}$, it reasonably
agrees with the restricted three-body results. In the bottom panel, the
capture probability is shown as a function of a scaled velocity  
$(m/m_T)^{1/6}(m_{par}/m_T)^{-1/2}v_\infty=(m_N/m_T)^{1/6}
(2\bar{I} Gm_T/a)^{1/2}$. Although the difference among them is small,  
if the parabolic approximation is used for all the cases, the solid and 
dashed lines should perfectly overlap with the black dashed-dotted line.

\begin{figure}[t!]
\plotone{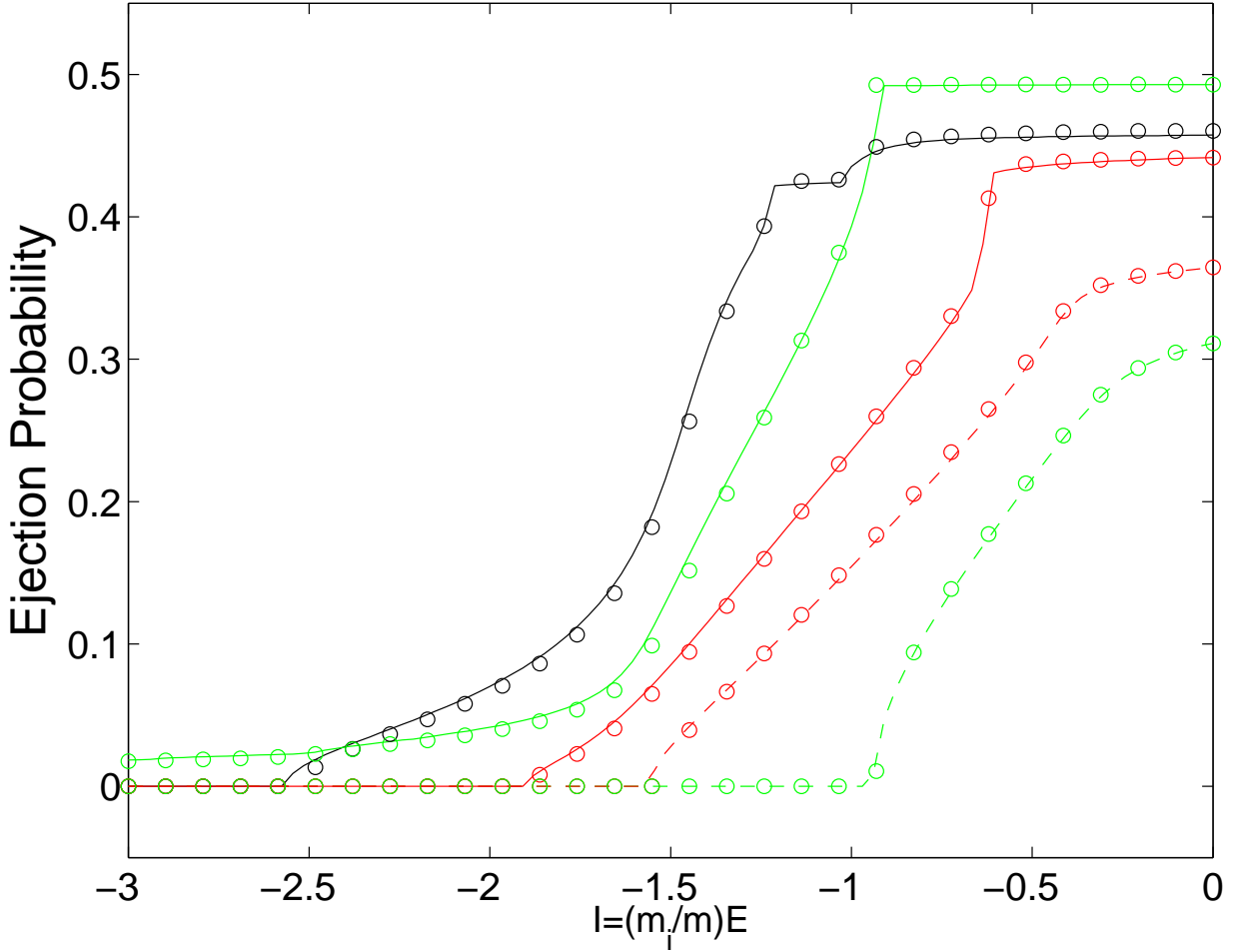}
\caption{
 Ejection probability as a function of 
 $\bar{I}=(m_i/m) \bar{E}$. The parabolic approximation is
 shown for $D=1$ (black line), $10^{-1}$ (green lines) and 
 $10^{-2}$ (red lines). The solid and dashed lines indicate 
 prograde orbit and retrograde orbit results, respectively.  
 The circles show the restricted  
 three body results for the corresponding cases with $M/m=10^6$ and 
 $m_i/m=1/4$. Energy is in units of $(Gm_1m_2/a)(M/m)^{1/3}$. 
\label{fig:peje}}
\end{figure}

\begin{figure}[b!]
\includegraphics[height=19cm]{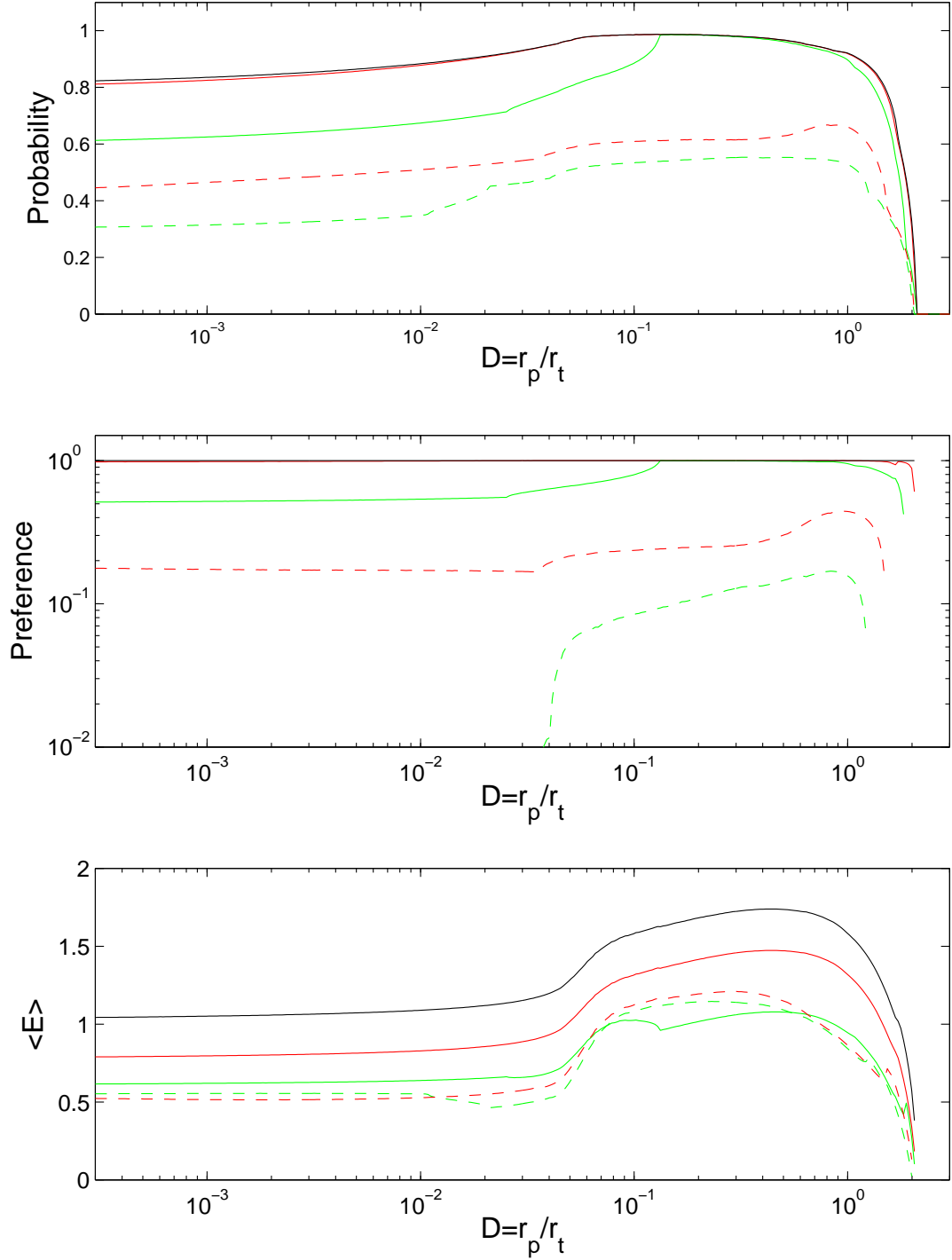}
\caption{
Ejection probability, preference and energy.
Parabolic orbit (black solid), 
$|r_a/r_t|=500$ (red solid), 200 (green solid), 130 (red dashed) 
and 100 (green dashed). The parabolic approximation is used. 
The ejection preference is the ratio of the ejection probabilities (the
 primary star/the secondary star). The ejection energy is evaluated
 by taking the phase average of all the ejected cases. 
Energy is in units of $(Gm_1m_2/a)(M/m)^{1/3}$.
 \label{fig:ave}}
\end{figure}

\begin{figure}
\includegraphics[height=19cm]{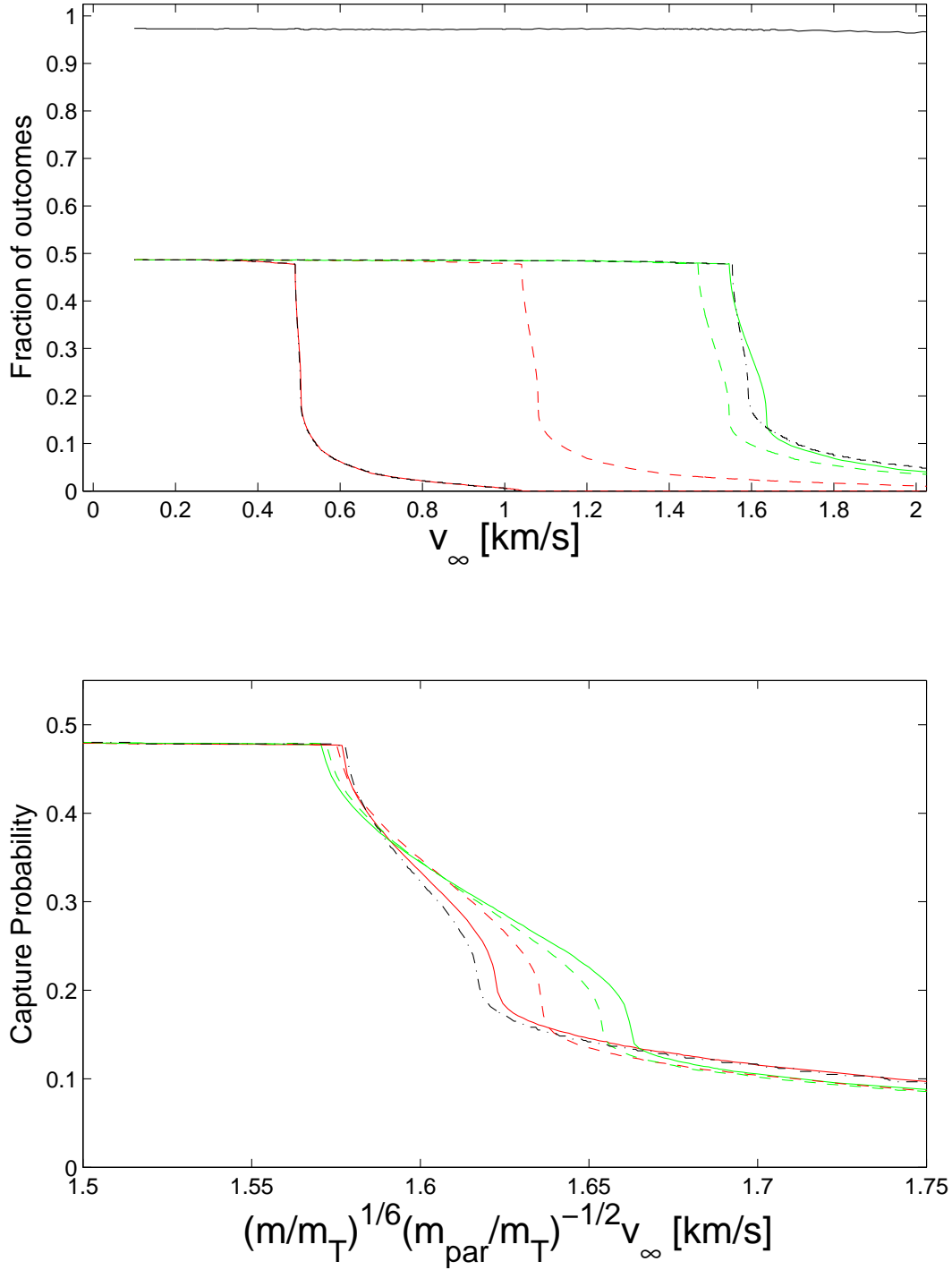}
\caption{Top panel: The restricted three-body approximation:
binary disruption chance (black solid),
capture chance for $m_2=0.1m_T$:
the primary (red solid) and the secondary (green solid),
capture chance for $m_2=m_T/2$:
the primary  (red dashed) and the secondary (green dashed).
The parabolic approximation for $m_2=0.1m_T$ (black dashed-dotted). 
Bottom panel: Capture chance as a function of the scaled 
velocity. $D\sim 0.45$ is assumed.
\label{fig:triton}}
\end{figure}

At the disruption, each member is captured with 
$50\%$ probability if the initial energy $\bar{I}$ is close to 
zero. Using the relation $P_{cap}(\bar{I})=P_{eje}(-\bar{I})$, figure
\ref{fig:peje} also shows that at high energies the capture is very rare. 
A transition occurs at an intermediate energy as shown in figure \ref{fig:peje}.
For $D=0.45$ prograde orbits, the capture probability sharply drops
from $\sim 50\%$ around  $\bar{I} \sim 1.4$. The critical value 
of $v_\infty$ at
which the capture probability of a binary member $m_1$ drops from $\sim
50\%$ is given by 
\begin{equation}
v_{\infty, ~crit} =\sqrt{\frac{2\bar{I} Gm_2}{a}}\paren{\frac{M}{m}}^{1/6}
\sim 0.49
\paren{\frac{\bar{I}}{1.4}}^{1/2}
\paren{\frac{m_2}{0.1m_T}}^{1/2} \paren{\frac{a}{20R_T}}^{-1/2}
\paren{\frac{M/m}{4.37\times10^3}}^{1/6} 
~\mbox{km/s}
\label{vcritical}
\end{equation}
This estimate well explains the critical velocities in figure
\ref{fig:triton}. If Triton has a heavier companion and/or a member of a
harder binary, the critical velocity could be higher, provided the
binary is disrupted by Neptune: $D \lsim 1$. The semi-major axis of the
captured member $m_1$ is 
\begin{eqnarray}
r_a&=&
 \frac{a}{2|\bar{E}_1|}\paren{\frac{M}{m_2}}\paren{\frac{m}{M}}^{1/3}
   \sim 1.6\times10^3R_N ~|\bar{E}_1|^{-1} \paren{\frac{a}{20R_T}}
   \paren{\frac{M/m_2}{4.80\times10^4}} \paren{\frac{M/m}{4.37\times10^3}}^{-1/3}
\label{tritonra}
\end{eqnarray}
where $|\bar{E}_1|$ is about $|\Delta \bar{E}|\sim 1$.
After capture, the orbit of Triton needs to shrink to the current
observed $r_a\sim14R_N$ and $e\sim10^{-5}$ either through tides or
other means (Correia 2009; Nogueira et al. 2011). The critical velocity 
and the semi-major axis for the secondary $m_2$ are obtained by
exchanging the subscript 1 and 2 in eqs (\ref{vcritical}) and (\ref{tritonra}).

Although a detailed modeling of the Solar system is beyond the scope of
this paper, the Sun can not be generally ignored in the capture
process (e.g. Philpott et al. 2010 Gaspar et al. 2011). 
Since the tidal radius $r_t \sim 18 R_N (m/m_T)^{-1}(a/20R_T)$ is much 
smaller than the radius of Neptune's Hill sphere $r_H \sim 4700 R_N$, 
Neptune completely dominates the attraction of a binary during the tidal 
encounter. Tidal effects from the Sun are negligible for the 
disruption process itself, which is the main focus of the paper. 
However, the permanent capture of Triton requires an additional 
condition that after the tidal breakup a capture orbit
should be well inside the Hill sphere.  
Detailed numerical simulations show that 
retrograde satellites (those orbiting in the opposite sense
as Neptune orbits the Sun  
\footnote{In the three-body encounter discussion, prograde motion
means that the binary center of mass is orbiting around a massive object
in the same sense as the binary members rotate around their center 
of mass. Then, a moon can be captured in a retrograde orbit around 
Neptune (the orbit is in the direction opposite to the rotation of 
Neptune) after the tidal breakup of a prograde binary 
(the angular momentum of the binary around Neptune and of a binary
member around the binary center of mass are aligned).}) of Neptune 
are more stable than prograde ones, and they are stable to distance of
$r_{stable} \sim 0.4r_H$ (Nesvorn\'y et al. 2003; Holman et al. 2004). 
Capture orbits are more eccentric than assumed in the numerical simulations, 
the stability region might be slightly smaller.  Considering that the
apocenter distance is about twice the semi-major axis (\ref{tritonra})
for highly eccentric orbits, the binary which had included Triton as a
member should satisfy a relation, 
\begin{equation}
\paren{\frac{a}{20 r_T}} \paren{\frac{m_2}{0.1 m_T}}^{-1} 
\paren{\frac{m}{1.1 m_T}}^{1/3} \lsim  0.4 \paren{\frac{r_{stable}}{0.3r_H}},
\end{equation}
where we have used $|\Delta \bar{E}| \sim 1$. 
A smaller value of the ratio $a/m_2$ gives a smaller apocenter distance and 
a higher critical velocity  (\ref{vcritical}). As shown in the top panel
of figure \ref{fig:D-dep} (the black solid and dashed lines), 
the phase-averaged value $\angle{|\Delta \bar{E}|}$ is slightly larger
than unity for prograde binaries, and slightly smaller for retrograde
binaries when the disruption probability is high. For a given encounter
velocity, a retrograde orbit of the binary center of mass around Neptune
with a prograde binary rotation is optimal to produce a stable capture
orbit.  

The ellipticity $\epsilon_N\sim 1.7\times10^{-2}$ of Neptune induces a
small deviation in its own gravitational potential from the point-mass
estimate especially in a non-planar configuration. However, the
deviation in the attraction force $\Delta F/F$ and the tidal
acceleration  $\Delta a_t/a_t$ is of order of 
$(R_N/r_t)^2\epsilon_N \sim 5\times10^{-5}$ at the tidal radius. The
effect is clearly negligible.

\section{Conclusions}
We have discussed how the members of a binary are ejected or captured
after a tidal encounter with a massive object. We have shown that the
ejection and capture dynamics can be well understood in the framework of
the restricted three-body approximation.  When the three-body exchange 
reaction happens, the orbit of the center of mass of the binary should be
almost parabolic  $|1-e| \lsim D (m/M)^{1/3}(m_{par}/m) \ll 1$ or
equivalently the semi-major axis is large $|r_a/r_t|\gsim (M/m)^{1/3}
(m/m_{par})$. The essential quantity to characterize the disruption process
is the energy change $\Delta \bar{E}$ at the encounter, which
practically depends only on three parameters: the penetration factor 
$D=r_p/r_t$, the binary rotation direction, and the binary phase $\phi$.
Except for the phase shift of $\pi$, the energy change is exactly the
same for the two members with arbitrary mass ratio.

In principle, for any positive (negative) orbital energy $E$ of the 
center of mass of a binary, the heavier member has more chance to be
ejected (captured), because it carries a larger fraction of the orbital
energy. However, if the orbital energy is close to zero, the difference
between their ejection (capture) probabilities 
becomes small, and there is practically no
ejection and capture preference. For a parabolic orbit, each member 
is ejected in exactly $50\%$ of the cases. 
The preference becomes significant when the absolute value of the
energy $|E|$ is comparable to the typical  
energy change $(Gm_1m_2/a)(M/m)^{1/3}$. On the other hand, 
if $|E|$ is much larger than the typical energy change, they are both 
ejected for $E>0$ or captured for $E<0$, and there is
no ejection or capture preference. 

Corresponding to the typical energy
change, we can define a critical encounter velocity  
$v_{\infty, ~crit}=(2 \bar{I} Gm_{par}/a)^{1/2}(M/m)^{1/6}$
for the capture process where the critical initial energy 
$\bar{I}\sim 0.5-1.5$ mainly depends on the penetration factor and the
orientation of the binary. Since the distribution of the energy change
is the same for the two members, the secondary star has a higher
critical velocity.  The tidal encounter might be
responsible for the capture of not only Triton but also other irregular
satellites in the Solar system (Morbidelli 2006, but see also
Vokrouhlick\'y et al. 2008; Philpott et al. 2010. The stability of
capture orbits for the solar perturbations also should be 
taken into account).  
However, the other irregular satellites are much lighter than 
Triton. The capture mechanism  
$v_{\infty,~crit} \propto m_{par}^{1/2}$ requires that these irregular
satellites have been in a binary with a very massive partner 
and that the less massive member has been predominantly captured. 

If a binary has a large orbital energy $E\gg (M/m)^{2/3} (Gm^2/a)$,
the disruption would happen slightly inside the usual tidal radius 
at which the BH tidal forces become comparable to the binary mutual
gravity forces. The abrupt disruption approximation provides a good
estimate of the energy change, and the typical energy change in the 
high energy regime is much larger than in the low energy regime. 
However, the change is not significant compared to the initial energy
$E$, the tidal encounter is not an efficient acceleration process anymore.
In the high energy regime, both members are ejected after the BH 
encounter.

In recent years, observations have identified a remarkable number of
hypervelocity stars (Brown et al. 2009; Tillich et
al. 2011). The distribution of line-of-sight velocities of hypervelocity
star candidates shows a long tail in the high velocity region ($v>275$
km/s) which includes comparable numbers of unbound and bound stars
assuming that the escape velocity of the Galaxy at 50kpc is 
$\sim 360$ km/s (Brown et al. 2009; Kenyon et al. 2008). Since the 
initial orbital energy is negligible when the three-body exchange 
reaction happens, the velocity distribution should reflect the 
distribution of the energy change $\Delta E$ at the tidal encounter 
(and the Galactic potential) with its dependence on the periapsis
distance, the binary orientation and phase. 

Binaries are supplied to the BH at 
the Galactic center predominantly from its radius of influence
$r_h \sim$ a few pc or even beyond it (e.g. Perets et al. 2007; 
another interesting possibility is that they might arise from the
nuclear stellar disk.  e.g. Madigan et al. 2011). Those that come 
from about the radius of influence are on elliptical 
orbits. Since the radius of influence is much larger than the tidal 
radius, the preferential ejection for high and low mass members 
$(m_1>m_2)$ is therefore irrelevant for stellar binaries 
($m_1/m_2\lsim 10$) if the binary semi-major axis is smaller than
$\sim$1AU. However,  stars with planets ($m_1/m_2\gsim 10^3$)
of semi-major axis of 1AU will predominantly eject the planets. 

We thank the referee, Hagai Perets, for his constructive comments, 
and Mostafa Ahmadi, Phil Armitage, Witold Maciejewski, and Ruth Murray
Clay for useful discussion.
This research was supported by STFC, ERC, and IRG grants, and Packard, 
Guggenheim and Radcliffe fellowships.


\newpage
\end{document}